**Density scaling of the diffusion coefficient at various pressures in viscous liquids**

*[accepted for publication in Phys. Rev. E (2009)]*

**Anthony N. Papathanassiou**

*University of Athens, Physics Department, Solid State Physics Section,*
*Panepistimiopolis, 157 84 Zografos, Greece*
*E-mail address: antpapa@phys.uoa.gr*

Fundamental thermodynamics and an earlier elastic solid-state point defect model [P. Varotsos and K. Alexopoulos, Phys. Rev B **15**, 4111 (1977); **18**, 2683 (1978)] are employed to formulate an analytical second-order polynomial function describing the density scaling of the diffusion coefficient in viscous liquids. The function parameters are merely determined by the scaling exponent, which is directly connected with the Grüneisen constant. Density scaling diffusion coefficient isotherms obtained at different pressures collapse on a unique master curve, in agreement with recent computer simulation results of Lennard-Jones viscous liquids, [D. Coslovich and C.M. Roland, J. Phys. Chem. B **112**, 1329 (2008)].

**PACS Indexes**: 64.70.P-; 61.72.J-; 62.20.de ; 66.10.cg

Viscous liquids are briefly characterized as 'solids that flow' rather than ordinary less viscous liquids [1,2] and exhibit many interesting features [3,4] and universal behaviour, which is not yet well understood. In the extreme viscosity limit (i.e., close to the calorimetric glass-transition) molecular transport is retarded and most molecular motion is vibrational [1] and the viscous liquid resembles a disordered solid [4]. Dynamics are strongly non-Arrhenius and the activation energy is strongly temperature-dependent for fragile glass formers [1]. A dynamic quantity $\chi$, such as structural relaxation time $\tau$, viscosity $\eta$ or diffusion coefficient D in viscous liquids is assumed to scale with density $\rho$ and temperature as:

$$\chi = F(\rho^\gamma / T) \qquad (1)$$

where $\rho$ denotes the density, $\gamma$ is a scaling exponent, T is the temperature and F is a function, which is a priori unknown [5]. The correlation of the exponent $\gamma$ with microscopic or thermodynamic quantities remains a matter of investigation. Computer simulations of Lennard-Jones liquids, with the exponent of the repulsive term taking the values 8, 12, 24 and 36, revealed that density scaling is valid and the exponent $\gamma$ is roughly one third of the exponent of the effective inverse power repulsive term [6]. Molecular dynamics also showed that strong virial/potential-energy correlations also reflect the effective inverse power law [7] and scaling occurs in strongly correlating viscous liquids [8]. On the other hand, following the Avramov entropy model [9] for the structural relaxation time, $\gamma$ was identified to the thermodynamic Grüneisen parameter $\gamma_G$ [10-12]. A series of interesting articles reviewing the peculiar properties of glass forming liquids were published recently [1,2,4,13].

Solid-state elastic models seem to play a prominent role in describing these phenomena. The distinctive role of thermodynamic point defect models for understanding the viscous state was mentioned recently by Varotsos [14]. In the present work, we start from thermodynamic definitions and by using elastic point defect models, we provide an analytical equation governing the density scaling of the diffusion coefficient in viscous liquids. The morphology of the scaling function agrees with up to date experimental results and computer simulations. The present formulation predicts that the scaling function is pressure insensitive, in agreement with recent computer simulations of binary Lennard-Jones systems, for various exponent values of the repulsive term of the potential results [6].

Isotherms of the logarithm of the relaxation time of viscous liquids as a function of pressure have a clear non-linear behavior [13, 15-17]. The pressure dependence of logarithm of the diffusion coefficient provided through molecular dynamics simulations [18] deviates from linearity, as well. $\ln D$ vs pressure shows a downward curvature. The increase of the (absolute) value of slope of the latter curve with pressure was speculatively interpreted, as a change in the transport mechanism in viscous liquids, occurring at pressure where hopping of particles become noticeable [18]. Alternatively, it was attributed [18], according to the free-volume theory, to a random close packing occurring at elevated pressure. However, the curvature in diffusion plots was thermodynamically interpreted earlier: Varotsos and Alexopoulos have proposed a generalized description of diffusion vs pressure isotherms [19], which can be used to analyze both linear and curved diffusion plots. If $g^{act}$ denotes the Gibbs free energy for diffusion, the corresponding activation volume is defined as $\upsilon^{act} \equiv \left(\partial g^{act}/\partial P\right)_T$. Since there is no physical argument to regard $\upsilon^{act}$ as constant, the compressibility of the activation volume may be defined as $\kappa_T^{act} \equiv -\left(\partial \ln \upsilon^{act}/\partial P\right)_T$ [19]; it can be positive, negative or zero. The data reported un Ref. 18 indicate that $\chi_T^{act} < 0$ for viscous liquids. The isothermal pressure evolution of the reduced diffusion coefficient is [19]:

$$\ln D(P) = -\left[\frac{\upsilon^{act}(0)}{kT} - \frac{\gamma_G}{B}\right]P + \left[\frac{\upsilon^{act}(0)\kappa_T^{act}}{2kT}\right]P^2 \qquad (2)$$

where $\upsilon^{act}(0)$ denotes the activation volume value *at zero (ambient) pressure*. It is evident that, whenever $\chi_T^{act}$ is zero (or $\upsilon^{act}$ is constant), Eq. (2) reduces to a simple well-known linear relation. From another viewpoint, the curvature may be interpreted if $\upsilon^{act}$ is not single-valued, but obeys a normal distribution [20, 21] Note that the quantity D appearing in Eq. (2) is a reduced one, with respect to the zero-pressure diffusion coefficient.

Starting from the definition of the isothermal bulk modulus $B \equiv -\left(\partial P/\partial \ln V\right)_T$, and recalling that $\rho \equiv m/V$ we get $B = \left(\partial P/\partial \ln \rho\right)_T$, or $B \equiv \gamma\left(\partial P/\partial \ln \rho^\gamma\right)_T$, where $\gamma$ is the scaling constant. Using the symbol $\rho$ for the reduced density, we get $\rho^\gamma = \exp\left(\frac{\gamma}{B}P\right)$. The later equation provides, to a first approximation, a linear relation between $\rho^\gamma$ and P.

$$\rho^\gamma \cong 1 + \frac{\gamma}{B}P \qquad (3)$$

For $P/B \leq 0.1$ and $\gamma=4$, the omission of higher order terms induces an error of less than 6%. For many viscous liquids, B is of the order of a few GPa [10, 22], so Eq. (3) works adequately well for pressure less than 1GPa, otherwise higher order terms are required. It is necessary to stress that the linear approximate relation between $\rho^\gamma$ and P is asserted so at to simplify the mathematical manipulation and *does not affect the underlying physics hidden behind the formulation*, which is the use of well-known solid state point defect models to describe the universal behavior of viscous state of condensed matter.

The interconnection of the scaling parameter γ with properties of viscous liquids is a matter of current interest. In Ref. [11], C. M. Roland *et al*, working on the scaling behavior of the structural relaxation time of super-cooled liquids, suggested that the scaling exponent γ is close to the value of the Grüneisen parameter, the exact relationship being model dependent. If the intermolecular potential is approximated by an inverse power law, various equations are derived, which correlate γ with $\gamma_G$. Describing the supercooled dynamics with an entropy model [10], $\gamma=\gamma_G$ is obtained. Following the latter visualization, by identifying the value of γ with $\gamma_G$, which is a measure of the anharmonicity of phonons, and assuming that the absolute value of the activation volume compressibility is comparable with the bulk compressibility (i.e., $\left|\kappa_T^{act}\right| \approx 1/B$) [23], Eqs. (2) and (3) combine to a unique relation:

$$\ln D(\rho^\gamma) = -\left\{\frac{\upsilon^{act}(0)}{kT} \cdot \frac{B}{\gamma} - 1\right\}(\rho^\gamma - 1) - \left\{\frac{\upsilon^{act}(0)}{2kT} \cdot \frac{B}{\gamma^2}\right\}(\rho^\gamma - 1)^2 \qquad (4)$$

Solid-state elastic point defect models suggest that the activation volume is proportional to the activation Gibbs free energy $g^{act}$ [24]. According to the cBΩ model [24-27], $\upsilon^{act} = B^{-1}\left(\frac{dB}{dP} - 1\right)g^{act}$. As explained in Ref. 28, the latter relation can take the form

$$\upsilon^{act} = \frac{2\gamma_G}{B} g^{act} \qquad (5)$$

In the viscous state, the activation enthalpy is *of the order* of 10kT (or a few tenths of kT) [2, 29, 30]. We can write $h^{act} \approx \Lambda kT$, where $\Lambda$ is a number of the order of 10, which is material dependent [30]. The activation entropy $s^{act}$ is only about a few k, thus, $g^{act} = h^{act} - Ts^{act}$ is of the same order of magnitude as $h^{act}$ is. Subsequently, Eq. (5) may rewritten as:

$$\upsilon^{act} \approx \Lambda \frac{2\gamma_G}{B} kT \qquad (6)$$

We note that we refer to a constant temperature (i.e., isotherms of diffusivity at various pressures) and, therefore, we skip the temperature dependence of the $h^{act}$ (fragility) and, subsequently, $g^{act}$. Eq. (6) is used to eliminate $\upsilon^{act}(0)$ from Eq. (4), which, recalling that $\gamma$ and $\gamma_G$ practically share the common value, reduces to:

$$\ln D(\rho^\gamma) \cong -\frac{\Lambda}{\gamma}(\rho^\gamma)^2 - \left\{2\Lambda\left(1-\frac{1}{\gamma}\right)\right\}\rho^\gamma + \left(2-\frac{1}{\gamma}\right)\Lambda - 1 \qquad (7)$$

We stress that Eq. (7) *does not simply result from a generalized diffusion equation by changing the independent variable from P to $\rho^\gamma$, but captures the interconnection of diffusion parameters with elastic properties of the material (within the frame of the cBΩ elastic solid state point defect model) and the universal feature of glass-formers that the activation enthalpy is of the order of 10kT (i.e., $h^{act} \approx \Lambda kT$, where $\Lambda$ is of the order of ten)*. Further work can improve the validity of Eq. (7): by including the temperature dependence of the activation enthalpy, which does it differently in different materials [30]. Moreover, additional correction terms may appear in Eq. (7) by considering second order (or higher) terms in Eq. (3). Simulations of Eq. (7), at constant temperature, are presented in Fig. 1. This equation predicts that:

(i) The (natural) logarithm of the reduced diffusion coefficient is a decreasing function of $\rho^\gamma$.

(ii) The function $\ln D(\rho^\gamma)$ is a second order polynomial with downward curvature. The latter form, which is based on physical arguments, is suitable to fit isothermal density scaling diffusion data, instead of using arbitrary equations [31].

(iii) The slope of the $\ln D(\rho^\gamma)$ curve depends on $\Lambda$, which is a characteristic of the material, and the scaling parameter $\gamma$, which is also an inherent characteristic of the viscous liquid and its value, according to the literature, is very close to the anharmonic Grüneisen constant [32].

(iv) Different $\ln D(\rho^\gamma)$ isotherms obtained at different pressures for the same viscous liquid, collapse on a unique master curve. This is due to the fact that $\Lambda$ and $\gamma$ are constant for the viscous liquid under study. The present formalism gives the theoretical interpretation of computer simulation results of Lennard-Jones liquids m-6 ($8 \leq m \leq 36$) in normal and moderately super-cooled states [6], which indicated that the diffusion coefficient plotted against $\rho^\gamma/T$ at different pressures, accumulate on a single curve [33].

The *density and temperature* scaling of dynamic properties of viscous liquids is relatively a recent speculation [34]. At present, apart from numerical simulations, experimental work on *density and temperature* scaling is available for the structural relaxation time and the viscosity, but missing for the diffusivity. At present only numerical results are available from important groups, which make predictions on the scaling of diffusivity [6,8]. Concerning the diffusivity, it seems that we are at a stage that simulations and theory are temporarily advancing in relation with the experimental work. The results of the present theoretical work can therefore compare with the available simulated experiments in Lennard-Jones liquids. The currently published simulations and the present theoretical work exhibit the emerging necessity of investigating experimentally the density and temperature scaling of diffusion coefficient in viscous liquids.

The extraction of Eq. (7), which was based on thermodynamic definitions and the cBΩ elastic solid-state point defect model, confirms the statement of Dyre [3] that viscous flow events can be correlated with defect motion in crystals: free energies from activation for self-diffusion are proportional to the isothermal bulk modulus (cBΩ model) and, if shear and bulk moduli are proportional to their temperature and pressure variation, then the cBΩ model becomes equivalent to the shoving model [3], which is based on the fact that activation energy is dominated by the work done to shove aside the surroundings [2, 35].

**Acknowledgements**

The author is grateful to Daniele Coslovich (Universita di Trieste, Italy) and Mike Roland (Naval Research Laboratory, USA) for making important suggestions.

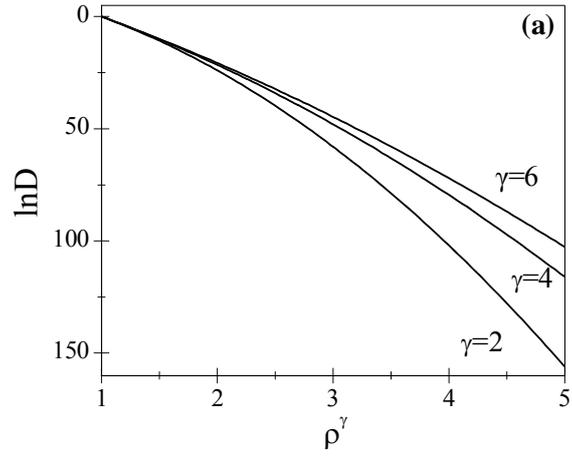

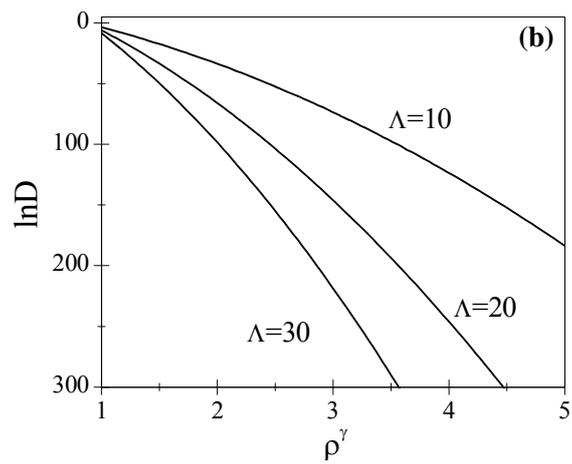

**Figure 1.** (a): Isothermal plots of lnD against $\rho^\gamma$, according to Eq. (7), considering $\Lambda=10$, for different values of the scaling exponent $\gamma$.

(b): Isotherms of lnD against $\rho^\gamma$, according to Eq. (7), for different values of the $\Lambda$ parameter and $\gamma=4$.

Note that D and $\rho$ are reduced dimensionless quantities.

appear in Eq. (7), which is practically negligible (i.e., when plotted on the same diagram, they superposition each other and can hardly be distinguished.)